\documentclass[twocolumn,amsmath, amssymb, superscriptaddress, pra]{revtex4}
\usepackage{graphicx}
\usepackage{subfigure}
\usepackage{dcolumn}
\usepackage{bm}
\usepackage{epsf}
\usepackage{amssymb}
\DeclareGraphicsExtensions{.eps}
\begin{document}

\title{\large\bf Topological Transitions and Bulk Wavefunctions in the SSH Model}

\author{David S. Simon}
\email[e-mail: ]{simond@bu.edu} \affiliation{Dept. of Physics and Astronomy, Stonehill College, 320 Washington Street, Easton, MA 02357} \affiliation{Dept. of
Electrical and Computer Engineering \& Photonics Center, Boston University, 8 Saint Mary's St., Boston, MA 02215, USA}
\author{Shuto Osawa}
\email[e-mail: ]{sosawa@bu.edu} \affiliation{Dept. of Electrical and Computer Engineering \& Photonics Center, Boston University, 8 Saint Mary's St., Boston, MA
02215, USA}
\author{Alexander V. Sergienko}
\email[e-mail: ]{alexserg@bu.edu} \affiliation{Dept. of Electrical and Computer Engineering \& Photonics Center, Boston University, 8 Saint Mary's St., Boston,
MA 02215, USA} \affiliation{Dept. of Physics, Boston University, 590 Commonwealth Ave., Boston, MA 02215, USA}

\begin{abstract} Working in the context of the Su-Schreiffer-Heeger (SSH) model, the effect of topological transitions on the structure and properties of
bulk position-space wavefunctions is studied for a particle undergoing a quantum walk in a one-dimensional lattice. In particular, we consider what happens when
the wavefunction reaches a boundary at which the Hamiltonian changes suddenly from one topological phase to another. An exact solution is constructed for the
wavefunction on both sides of the boundary. Under some conditions, it is found that the  probability of transition into the region of the
second, topologically distinct Hamiltonian is strongly suppressed. When the boundary is encountered, the wavefunctions tend to be strongly reflected, and by
appropriate choice of system parameters leakage into the second region can be made negligible. Therefore, it is possible to arrange a high degree of bulk
wavefunction localization within in each region. This ``topologically-assisted'' suppression of transitions, although not of direct topological origin itself, exists only
because of the presence of a change in the topological properties of the Hamiltonian. We give a quantitative examination of the reflection and transmission coefficients of
incident waves at the boundary between regions of different winding number.
\end{abstract}

\maketitle

\section{Introduction}\label{introsection}

\begin{figure}
\centering
\includegraphics[totalheight=1.1in]{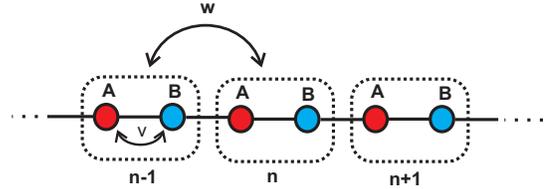}
\caption{The SSH Hamiltonian describes motion of a particle hopping on a chain of sites with two substates per site. $v$ and $w$ are respectively the intracell and intercell hopping amplitudes per unit time.}
\label{sshfig}
\end{figure}

The Su-Schreiffer-Heeger (SSH) model \cite{su} was originally proposed as a model of electron behavior in polymers, but has become widely used as a simple model
in which topological phase transitions may occur. The system is defined in terms of two parameters, $w$ and $v$, representing hopping amplitudes between two
distinct types of lattice sites. Each unit cell is formed by a pair of lattice sites, one of each type (Fig. \ref{sshfig}). As the quasimomentum $k$ varies
across the full Brillouin zone, the Hamiltonian $\hat H$ traces out a closed curve in a two-dimensional space. When the parameters obey $v>w$ this curve avoids
enclosing the origin, where $\hat H$ becomes singular, and so the Hamiltonian exhibits a vanishing winding number about the singularity
\cite{hasan,asboth,kitagawatop}. For $v<w$, the curve encloses the singular point, and has winding number $1$ about the origin. At the borderline case $v=w$, a
transition between two topological phases occurs, with the winding number making a discontinuous jump. Since the SSH lattice has two distinct sublattices,
corresponding to two ``internal states'' within each unit cell, the energies form two bands separated by a finite gap. However, when $v=w$, the gap disappears,
and at this point transitions between different winding numbers occur.

In treatments of the SSH model, the emphasis is usually on the Hamiltonian and the energy bands. Wavefunctions are normally given less attention, and when they
are discussed the focus is usually on Bloch states in momentum space or on the localized edge states that appear between regions of different winding number.
However, recently the behavior of particles undergoing quantum walks in SSH-like systems has become an important topic of research; in particular photonic
quantum walks in linear optical systems have been shown to simulate topological states of the same type that appear in SSH-like systems
\cite{broome,kit1,kit2,taras,kit3,cardano,sim3}. In these photonic systems, the particle is inserted at a fixed location, then at a later time (after some number
of discrete time steps) its final position distribution is measured. Thus, position-space wavefunctions in the bulk are of significant interest and hold the key
to a more complete understanding to transitions between topologically distinct regions.

\begin{figure}
\begin{center}
\subfigure[]{
\includegraphics[scale=.26]{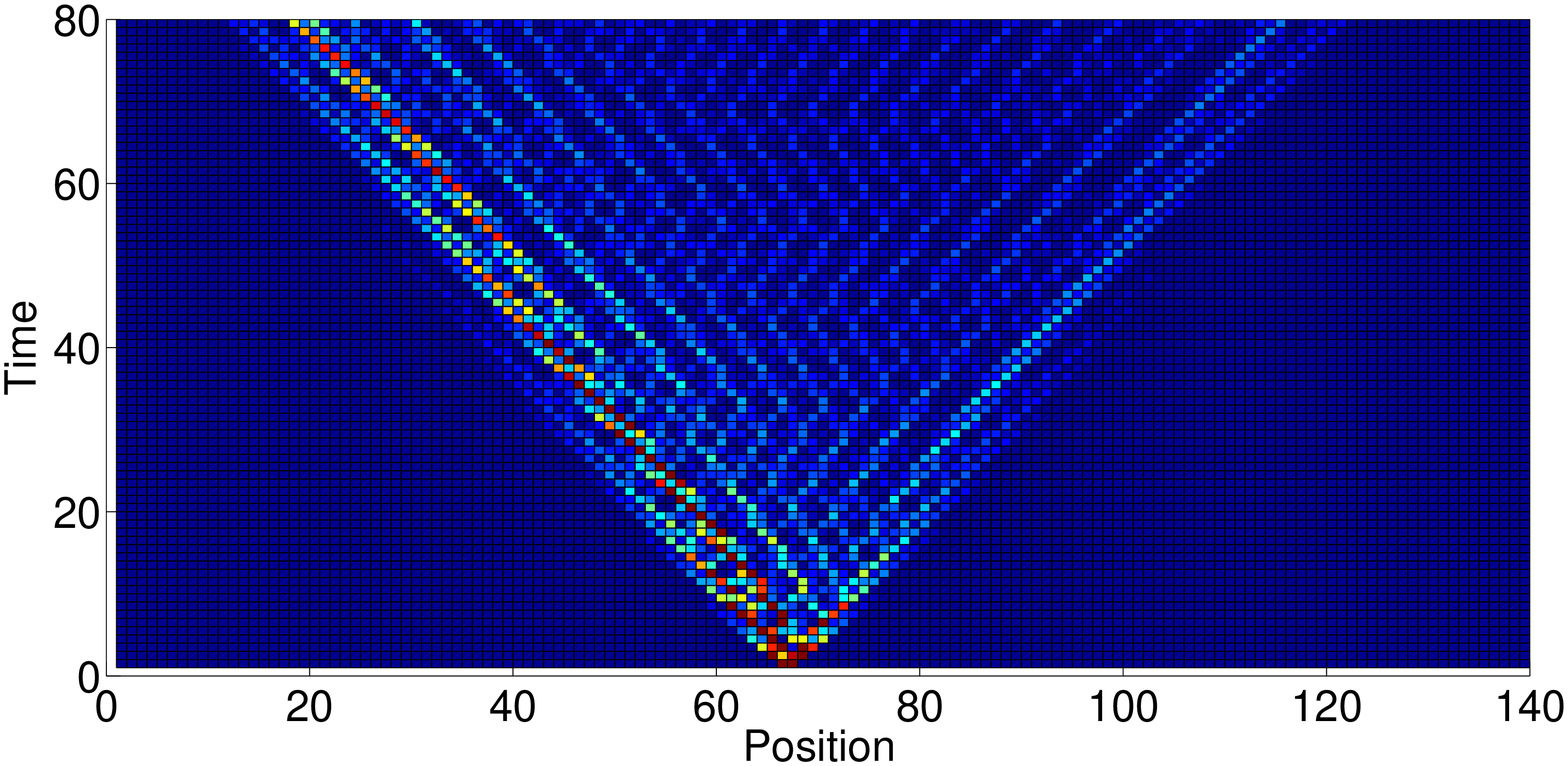}}
\qquad  \subfigure[]{
\includegraphics[scale=.26]{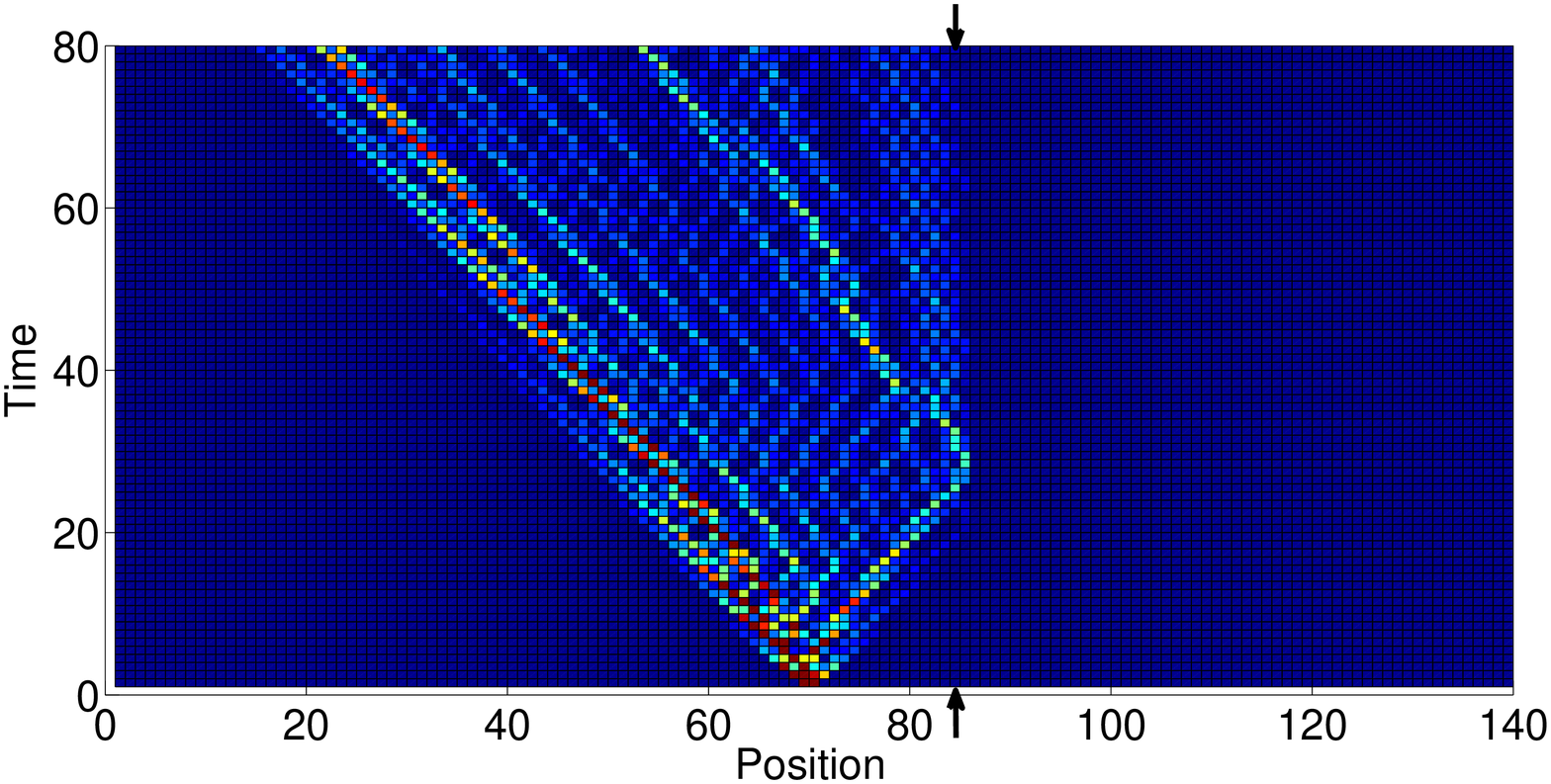}}
\caption{Calculated probability distribution of detecting a photon at position $x$ at time $t$ as the photon undergoes a one-dimensional quantum walk.
The system in which the photon is
walking is the linear optical arrangement of \cite{sim2}. (a) The Hamiltonian is the same (of winding number zero) throughout. The photon is inserted
into the system at position $x=70$ and then exhibits the ballistic evolution characteristic of quantum walks. (b) The same, except that
now the parameters of the system
change at position $x=85$ (marked by the arrows) so that the winding number of the Hamiltonian is $0$ to the left of that point and $1$ to the right. It can be seen that
propagation into the region of ``wrong'' winding number is strongly suppressed. }\label{onechainfig}
\end{center}
\end{figure}

Figure \ref{onechainfig} shows a simulation of a photonic quantum walk. In (a), the entire system has the same Hamiltonian, and the photon spreads ballistically
in both directions from the point of insertion, displaying the well-known probability distribution of quantum walk systems. However, in (b) the parameters of the
system abruptly change at lattice site $85$, causing the Hamiltonian on the right side of that point to have a different winding number than the Hamiltonian to
the left. It can be clearly seen that penetration of the photon into the region on the right is strongly suppressed, with some of the amplitude collecting at the
boundary and most of the rest reflecting back into the original region. Such a suppression of transitions into regions of different winding number may be seen in
experimental data as well (see Fig. 3 of \cite{kit3} for example), although in experiments the effect is somewhat obscured due to the small number of steps
measured and the presence of the localized state at the boundary that extends a few steps into the second region.  Possibly because of these obscuring factors,
this effect has not been much remarked upon. One possible interpretation is simply a mismatch between energy levels on the two sides of the boundary, but it is
also possible that \emph{the change in topology plays a role} regardless of the energies. Our goal here is to examine this question more closely.

More specifically, this paper looks in detail at position-space wavefunctions  of SSH systems and at what happens to them when the topology of the Hamiltonian changes. We
assume that the Hamiltonian depends on some parameter that is under the control of experimenters (for example, the vertex phase parameters in the quantum walk
systems of \cite{sim3,sim2} or the rotation angles of \cite{broome,kit1,kit2,taras,kit3}). We further assume that the parameter varies in such a way that
at some position $x$ it causes the winding number of the Hamiltonian to change. Although all of the considerations in the following sections apply
equally to other physical realizations, we will assume for the sake of specificity that the particles involved are photons.


At the boundary between the two regions there should be a transition between states that are asymptotically (far from the boundary) eigenstates of the original
Hamiltonian to eigenstates of the topologically-altered Hamiltonian. We show in the following that if the two Hamiltonians are of different winding number, then
such transitions are partially suppressed by an amount depending on the values of the $v$ and $w$ parameters of the two Hamiltonians,
with strong reflection at the interface. The net result is that in a system of regions governed by Hamiltonians of different winding number,  if the hopping
amplitudes are well-chosen, then states will strongly tend to remain in the region where they started and resist propagation into other regions. Although we have
been speaking of spatial regions here, the same will apply to different regions of some more abstract parameter space: for example if a system is arranged such
that photons see a polarization-dependent Hamiltonian that has different winding number for vertical and horizontal cases, then polarization flips will be
suppressed by the same mechanism. This has obvious applications, for example in reducing the likelihood of polarization-flip errors in optical information
processing systems. It will be shown elsewhere that Hamiltonians with such polarization-dependent winding numbers can be readily engineered using linear optics.

The existence of localized, topologically protected states at the boundaries between regions with different winding number is well-known. The results here imply
that under appropriate conditions there is also a measure of ``protection'' attached to the bulk wavefunctions, in the sense that propagation into spatial regions or
parameter regions with different topological properties is suppressed. The degree of transition suppression between these regions depends on the parameter values themselves, as well as the
presence or absence of discrete topology changes. In this paper we only examine the simplest case, in which the two hopping parameters are interchanged at the
boundary: the value of $v$ on the left equals the value of $w$ on the right, and vice-versa.  A quantitative study of how the suppression varies as the values of
$v$ and $w$ move away from the pure exchange case will be carried out elsewhere.

Note further that if one can strongly suppress transitions of the bulk wavefunction between regions of parameter space of different topology, one can associate
the wavefunction localized in a given region with a Hamiltonian of particular topology. Then linear combinations of states associated to different winding number Hamiltonians may be
formed, allowing winding number-based qubits. Such linear combinations can be easily arranged, for example, by inserting linear combinations of polarization states, with different polarizations being governed by Hamiltonians of different winding number.  These qubits can be thought of as being encoded into either the particle state or the associated Hamiltonian. Gates can then
be made that act by altering the Hamiltonian, with readout accomplished by making measurements on the states.

In the coming sections, we will see that the presence of a change in the topology of the Hamiltonian (a discrete change in its winding number) affects the
transmission and reflection coefficients at the boundary point. In the absence of the topology change, the transmission coefficient is uniformly equal to
$100\%$. However,  when the topology change is in place the transmission coefficient becomes a continuous function of the hopping parameters, and for some
parameter ranges the transmission can be made very small.

This discussion points out that there is interplay between discrete, topologically-based variables (winding number) and continuous variable (transmission
coefficient) that may be affected by them. Such interplay exists in many other contexts. Consider, for example, a particle striking a potential barrier of height
$V_0$ from the left. The barrier is discrete: you are either at the top (on the right) or at the bottom (on the left), with no possibility of being in between.
The top/bottom distinction only exists because of the discrete jump in potential at the origin. The transmission coefficient for the particle to move from left
to right, however, is continuous as a function of energy. Despite being continuous, it depends on the existence of the discrete potential step. When the particle
energy $E$ satisfies $E<V_0$ the transmission amplitude $t$ is uniformly zero, but as the energy increases to the point where
$E$ exceeds $V_0$, t becomes a continuously-varying function of $E$. So the value of the transmission amplitude is affected by the discrete variable, and yet remains a continuous function of its parameters
(energy in this case), and for some parameter values becomes negligible. The situation in our case is directly analogous, with the discrete potential step
replaced by a discrete topology change and the transmission amplitude's dependence on the energy replaced by dependence on hopping parameters. 	

A similar analogy occurs at the edge of a step-index optical fiber. The refractive index changes discretely, but that change affects the properties of the
reflected and evanescent waves. Without the discrete jump there would be no evanescent wave at all; but the properties of that wave (penetration depth, etc.)
still depend continuously on other parameters such as the angle of incidence.

In this paper, we show that reflections occur at the points of sudden topological changes, in the same manner that reflections occur at any other abrupt change such as a sudden change in potential energy, acoustic impedance, or refractive index.
We use the Su-Schreiffer-Heeger (SSH) system as an example system. In many applications of the SSH model, the electron or other hopping particle is
treated as a point particle, perfectly localized at a given lattice site at each moment. In reality, we know that quantum mechanical wavefunctions typically have
a finite spread to them and this spread needs to be taken into account to study quantities like scattering amplitudes and tunneling rates. In the current paper
we are looking at the transition rate from one side of a discrete boundary to the other side, and this will clearly be dependent on the wave-like properties of
the particle. So we focus on the position-space wavefunction and expand it in a convenient basis. In the context of a particle interacting with a discrete
lattice system, a convenient basis is the Wannier basis, in which the wavefunction is built out of basis states localized near each lattice site. A particle
initially localized near one site will exhibit a quantum walk \cite{kempe,amba,portugal,kitagawatop}, evolving into a superposition of states localized at many
sites; what we compute is the rate of transmission of this quantum walk state from one side of a topological boundary to the other.

The plan of the paper is as follows. In Section \ref{sshsection} we briefly review the SSH model and set up the notation for what follows. In Section
\ref{wavesection} we construct the position- and momentum-space wavefuntions expressed in the Wannier basis. In Section \ref{transition} we carry out a
quantitative examination of reflection and transmission of the wavefunction at the boundary, making use of the transmission coefficient calculated explicitly in
the appendix. Finally, we summarize the results and discuss further aspects of them in Section \ref{concludesection}.

\section{Brief Review of SSH Model}\label{sshsection}

The Su-Schreiffer-Heeger (SSH) Hamiltonian \cite{su} in one dimension describes the hopping of particles along the length of a bipartite lattice. A closely related model arose independently in quantum field
theory \cite{jackiw}.

The SSH system is shown schematically in Fig. \ref{sshfig}. There is a  lattice of unit cells, labeled by integer $n$, each of which contains two
subsites, denoted as $A$ and $B$; these subsites represent two possible ``internal'' states at cell $n$.  There is an amplitude per unit
time $v$ to switch between the two states within the same cell, and an amplitude per time $w$ to hop to an adjacent lattice site. Hopping to an adjacent  site is always accompanied by a change of the internal state. By redefining the basis states if necessary, the hopping amplitudes $v$ and $w$ can always be chosen, without loss of generality, to be real.

\begin{figure}[h!]
\centering
\includegraphics[totalheight=1.2in]{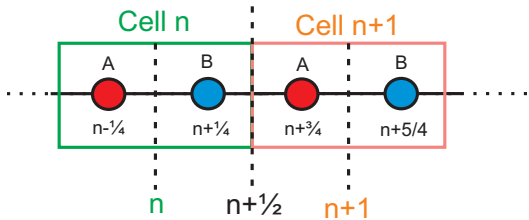}
\caption{Two unit cells of the lattice. The coordinates are chosen so that the center of each cell is at an integer-valued location. The two
subsites $A$ and $B$ are equally distant from the cell's center, so they are separated by half a unit, at locations
$n\pm {1\over 4}$, for $n=1,2,\dots ,N$}\label{coordfig}
\end{figure}

Let $R$ represent lattice positions and $r$ be the position of the particle moving through the lattice. It is convenient to take the center of each unit cell to
be at integer-valued positions, $R=n$, for $n=1,2,\dots ,N$, midway between the $A$ and $B$ subsites, as in Fig. \ref{coordfig}. So the spacing between cells is
one unit and the spacing between the $A$ and $B$ subsites within a cell is $1/2$ unit.  Then the $A$ subsites are at locations $n-{1\over 4}$ and the $B$ sites
are at $n+{1\over 4}$. In order to avoid edge effects, we may take periodic boundary conditions, so that site $n=N+1$ is identified with site $n=1$, or we may
simply take $N$ to be very large.

The position-space Hamiltonian is of the form:
\begin{eqnarray}\hat H &=& v \sum_{n=1}^N \left( |B,n\rangle \langle A,n | +  |A,n\rangle \langle B,n|  \right)\label{sshhamiltonian} \\
& & + w\sum_{n=1}^{N-1} \left( |A,n+1\rangle \langle B,n| +  |B, n\rangle \langle A,n+1|  \right) .\nonumber\end{eqnarray} Here, $|A,n\rangle$, for example, denotes the state with a particle at site $n$ in substate $A$.

At each fixed cell $n$ or each fixed momentum $k$, this Hamiltonian is therefore a two-dimensional matrix, and can be expanded in terms of the identity matrix
and the Pauli matrices; for example, in momentum space one may write
\begin{equation}\hat H(k) =
d_0(k) I +\bm d(k)\cdot \bm\sigma .\label{Handd}\end{equation}
This describes dynamics in a two dimensional ``internal'' subspace labeled by the two substates present at each lattice site. Generically, the two energy levels
are separated by a $k$-dependent gap.

The Hamiltonian is completely characterized by the 4-dimensional vector $\left\{d_0(k),\bm d(k)\right\}$. In the SSH model $d_0=d_z=0$, leaving $\bm d(k)$ confined to a plane. There is a singular
point in this plane, at $\bm d(k)=0$, where the phase of the Hamiltonian becomes indeterminate and the energy gap between bands vanishes.  As $k$ is varied
across a full Brillouin zone, $\bm d(k)$ traces out a closed curve. These curves can be divided into two distinct classes: those that encircle the singular point and those that don't. In other words, those whose winding number about the origin is $\nu =1$, and those of winding number $\nu =0$. The winding
number is highly stable in the sense that local perturbations causing continuous variations of the parameters cannot stimulate transitions between the discrete
topological classes.  Only a strong disturbance that alters the global structure of the system can cause the winding number to change.

In momentum space, the Hamiltonian is block diagonal, with blocks at each $k$ value of the form \begin{equation}\hat H(k) =\left(
\begin{array}{cc}0 & v+w\; e^{-ik} \\ v+w\; e^{+ik} & 0\end{array}\right)=\left(
\begin{array}{cc}0 & z \\  z^\ast & 0\end{array}\right) ,\end{equation} where $z=v+w\; e^{-ik}$. The two dimensions of this matrix correspond to the two subsites $A$ and $B$
inside the unit cell.

An alternative form of this Hamiltonian will be useful.
The off-diagonal terms can
be written in polar form in the complex
plane:  \begin{eqnarray}  z &=& v+w\; e^{-ik} \; =\; e^{-ik/2} \left( ve^{ik/2} +w e^{-ik/2} \right) \\
&=&  e^{-ik/2} \left( (v+w) cos\left({k\over 2}\right) +i(v-w) sin\left({k\over 2}\right) \right) \\
&=&  e^{-ik/2} \left( (v+w)^2 \cos^2\left({k\over 2}\right) \right. \\ & & \left. \qquad \qquad
+(v-w)^2 \sin^2\left({k\over 2}\right) \right)^{1/2} e^{i\theta_k}\nonumber \\
&=& E_k e^{i\theta_k-{ik/2}},\end{eqnarray} where
\begin{eqnarray}E_k &=& \left( (v+w)^2 \cos^2\left({k\over 2}\right) +(v-w)^2 \sin^2\left({k\over 2}\right) \right)^{1/2} \\ &=&
\left( v^2+w^2 +2vw\cos k \right)^{1/2} \end{eqnarray} is the absolute value of $z$, while
\begin{eqnarray}\theta_k &=& \tan^{-1}\left( {{Im(z)}\over Re(z)}\right) \\  &=& \tan^{-1} \left({{(v-w)}\over {(v+w)}}\tan {k\over 2}\right)
\label{tantheta}\end{eqnarray} is the phase.
So the Hamiltonian can be written as
\begin{equation}H(k)=E_k\left(
\begin{array}{cc}0 & e^{i\theta_k-ik/2} \\ e^{-i\theta_k+ik/2} & 0\end{array}\right) ,\label{hampolar}
\end{equation} showing clearly the winding of $H$ in the complex plane as the angle $\theta_k$ changes.
The factor $\theta_k-{k\over 2}$ in the exponent is, up to a constant, a geometric Berry phase.

The eigenvalues are \begin{equation}E_\pm (k)=\pm E_k =\pm \sqrt{v^2+w^2 +2vw\cos k} .\end{equation} It is then easy to verify that the eigenvectors are (up to
an arbitrary overall phase):
\begin{equation}|\pm\rangle = {1\over \sqrt{2}}\left( \begin{array}{c} 1 \\ \pm e^{-i(\theta_k -{k\over 2})}\end{array}\right) \equiv  {1\over \sqrt{2}}
\left( \begin{array}{c} u_\pm \\ l_\pm\end{array}\right).\label{lpmupm}\end{equation} The upper component corresponds to the $A$ substate, the lower to the $B$ substate.

It should be kept in mind that the quasi-momentum $k$ and the quasi-energy $E$ are only defined modulo $2\pi$.

\section{Position-space Wavefunctions}\label{wavesection}

Ignoring for the moment the $A$ and $B$ subcells, consider a simple lattice with $N$ sites at positions $R=1,2,\dots ,N$. One may construct the one-dimensional Bloch
wavefunctions for particle propagating through the lattice, \begin{equation}\psi_k(r)=e^{ikr}u_k(r),\label{psir}\end{equation} where $r$ is the particle
position, and the allowed momenta are
\begin{equation}k_n=  {{2\pi n}\over N}-\pi  ,\label{kvalues}\end{equation} with $n=1,\dots , N$. We take the first Brillouin
zone to run over the interval $-\pi<k\le +\pi$. For a single $k$ value, the corresponding position space wavefunction can be written in terms of the Wannier
functions $\phi_R(r)=\phi(r-R)$ \cite{wan,kohn,cloiz}:
\begin{eqnarray}u_k(r)&=&{1\over \sqrt{ N}} \sum_R e^{-ik(r-R)}\phi(r-R) \\ \psi_k(r) &=& e^{ikr}u_k(r)\; =\; {1\over \sqrt{ N}} \sum_R e^{ikR}\phi(r-R) .
\end{eqnarray} Wannier functions are widely used in solid state physics and other areas, and are defined as the Fourier transforms of the Bloch wavefunctions
with respect to the discrete lattice positions, \begin{equation}\phi_R(r) = \phi(r-R) ={1\over \sqrt{N}}\sum_k e^{-ikR}\psi_k(r).\end{equation} There is one such
function for each point in the crystal lattice and they are strongly localized near those lattice sites.

The functions centered at different lattice sites are orthogonal, \begin{equation}\int\phi^\ast (r-R) \phi (r-R^\prime ) dr =\delta (R-R^\prime ) ,\end{equation}
so the Wannier functions form a complete basis for spatial wavefunctions on the lattice. For the SSH model, these get multiplied by a two-dimensional column
matrix in the internal $A/B$ space spanned by the two subsites at each $n$.

Now introduce the $A$ and $B$ sublattices, so that the lattice sites are shifted to $R=n\pm {1\over 4}$, where $n=1,2,\dots ,N$. This splits each term in the sum of
Eq. \ref{psir} into two terms, one shifted the left by ${1\over 4}$ (the $A$ terms) and the others shifting to the right by the same amount (the $B$ terms), as in Fig. \ref{coordfig}.
Because there are two substates at each unit cell, the energy splits into two bands, as in section \ref{sshsection}.  Taking into account that the $l_\pm$ and $u_\pm$ defined in Eq. \ref{lpmupm} gain phases $\pm k/4$ from the shifts away from the cell center, the state then becomes:
\begin{eqnarray}\psi_{k\pm}(r) &=& {1\over \sqrt{2 N}} \sum_{n=1}^N \left[ \left( u_\pm e^{ik/4} \right) e^{ik(n-{1\over 4})}\phi\left( r-\left( n-{1\over
4}\right)\right) \right. \nonumber\\ & & \; \left.
+ \left( l_\pm e^{-ik/4} \right) e^{ik(n+{1\over 4})}\phi\left( r-\left( n+{1\over 4}\right)\right) \right]  \\
&=& {1\over \sqrt{2 N}}\sum_{n=1}^N e^{ikn}\left[  \phi\left( r- n+{1\over 4}\right) \right. \nonumber\\ & & \qquad\qquad \left.
\pm e^{-i\theta_k+ik/2}\phi \left(r- n-{1\over 4}\right) \right] ,\label{kpm2}
\end{eqnarray} where the $\pm$ labels correspond to the two eigenstates in the upper ($+$) and lower ($-$) bands. The first Wannier function inside the square bracket is centered at the $A$ subsites, while the second function is localized near the $B$ subsites.

Finally, the main interest here is not in wavefunctions of fixed $k$, but rather in states that are initially localized in position. Any position-space wavefunction at $t=0$ can be
expanded in terms of energy eigenstates,
\begin{equation}\psi (r)=\sum_{k=1}^N\left(  A_{k+}\psi_{k+}\left( r\right)+ A_{k-}\psi_{k-}\left( r\right)\right) .\end{equation}
The $A_{k\pm} $ coefficients can be found by taking the overlaps between  $\psi (r) $ and the known initial wavefunction.   There are two possibilities for the initial state: the photon may be inserted at an $A$ subsite or a $B$ subsite.  In the first case, we take the initial state to be \begin{equation}\psi_A(r)=\phi \left(r-n_0+{1\over 4}\right),\end{equation} with $n_0$ being the label of the initial lattice site. In the latter case, we take the wavefunction to be \begin{equation}\psi_B(r)=\phi \left(r-n_0-{1\over 4}\right),\end{equation}.

Making use of the orthonormality of the Wannier functions, it is straightforward then to find that the wavefunctions at $t=0$ for the two cases are
\begin{eqnarray}
\psi_{A}(r)&=& {1\over \sqrt{2N}}\sum_k e^{-ikn_0} \left( \psi_{k+}\left( r\right) +\psi_{k-}\left( r\right)\right)\\
\psi_{B}(r)&=& {1\over \sqrt{2N}}\sum_k e^{-ikn_0} e^{+i(\theta_k-{k\over 2})}\nonumber \\
& & \qquad \qquad \times \left( \psi_{k+}(r)-\psi_{k-}(r)\right) .
\end{eqnarray}
Clearly, the $k$ values are uniformly distributed in probability, as would be expected for a wavefunction localized in space.



Again using the orthonormality of the Wannier functions, it follows readily that these initially-localized functions also form an orthonormal set:

\begin{eqnarray} \int \psi^\ast _A(r) \psi^\prime_A(r)\; dr &=&   \delta (n_0^\prime -n_0) \\
 \int \psi^\ast _B(r) \psi^\prime_B(r)\; dr  &=&   \delta (n_0^\prime -n_0) \\
\int \psi^\ast _B(r) \psi^\prime_A(r)\; dr &=& 0.
\end{eqnarray} where $n_0$ and $n_0$ are the initial cells of the two wavefunctions.

All of the expressions above were at $t=0$.  Evolving forward in time, the energy eigenstates become
\begin{eqnarray} \psi_{k+}(r,t) &=& e^{-iE_kt} \psi_{k+}(r) \\ &=&
{1\over \sqrt{2 N}}\sum_{n=1}^N e^{i\left( kn-E_kt\right)}\left[  \phi\left( r- n+{1\over 4}\right) \right. \nonumber\\ & & \quad\qquad \left.
+ e^{-i\theta_k+ ik/2}\phi \left(r- n-{1\over 4}\right) \right] \label{k+1}\\
\psi_{k-}(r,t) &=& e^{+iE_kt} \psi_{k-}(r) \\ &=&
{1\over \sqrt{2 N}}\sum_{n=1}^N e^{i\left( kn+E_kt\right)}\left[  \phi\left( r- n+{1\over 4}\right) \right. \nonumber\\ & & \quad\qquad \left.
- e^{-i\theta_k+ ik/2}\phi \left(r- n-{1\over 4}\right) \right]  \label{k-1}
\end{eqnarray}
This means that the $A$- and $B$-type wavefunctions become
\begin{eqnarray} \psi_{A}(r,t)&=&  {1\over N}\sum_{kn} e^{ik(n-n_0)} \left( \phi \left( r- n+{1\over 4}\right)   \cos (E_kt)\right. \nonumber\\ & & +
\left. ie^{-i\theta_k+{i{k\over 2}}} \phi \left( r- n-{1\over 4}\right)   \sin (E_kt)\right) \nonumber \\
\psi_{B}(r,t)&=&  {1\over N}\sum_{kn} e^{ik(n-n_0)}\label{psiAA}\\ & & \times \left( i e^{i\theta_k-{i{k\over 2}}}\phi \left( r- n+{1\over 4}\right) \sin (E_kt)\right.
\nonumber\\ & &  \qquad +
\left.  \phi \left( r- n-{1\over 4}\right)   \cos (E_kt)\right)\label{psiBB}
\end{eqnarray}  where $t$ is an integer multiple of some discrete time interval $T$.
Time evolution only alters the phase of each $k$ component by  a factor $e^{iE(k)t}$, and therefore the probability of finding each $k$
value is constant in time. However interference between different terms in the sum leads to
nontrivial time evolution for the spatial distribution.


Note that the factor of $ e^{i\left( kn-E_kt\right)}$ in Eq. \ref{k+1}, which implies that the states $\psi_{k+}$ on the positive-energy band are right-moving
for positive $k$ and left-moving for negative $k$. The negative-energy states $\psi_{k-}$ move in the opposite direction: left for $k>0$ and right for $k<0$.

\section{Topological transitions}\label{transition}

Recall that the topological sector of the system is determined by whether $v>w$ or $v<w$. Consider two sets of values of $(v,w)$ and $(v^\prime ,w^\prime )$,
leading to two values of the phases, $\theta_k$ and $\theta_k^\prime $, and corresponding wavefunctions $\psi_{k\pm} (r)$ and $\psi_{k\pm}^\prime (r)$. These
wavefunctions are eigenstates of Hamiltonians $\hat H (k)$ and $\hat H^\prime (k)$. We take the full Hamiltonian of the system to be $H(k)$ for $n=-N+1,\dots ,0$
and $H^\prime (k)$ for $n=1,\dots ,N$, where $N$ is assumed large enough to ignore effects from the ends.

Now suppose that $\hat H (k)$ and $\hat H^\prime (k)$ differ in winding number. Among other things, this implies that the signs of $\theta_k$ and
$\theta_k^\prime$ are opposite at the same value of $k$. We will restrict ourselves to the simplest case and assume that the initial and final hopping
amplitudess are simply interchanged: $v^\prime =w$ ands $w^\prime =v$. In this case, we find that: \begin{equation}\theta_k^\prime =-\theta_k
=+\theta_{-k}.\end{equation}

Referring to  Eq. \ref{kpm2} and suppressing the time-dependent factors for simplicity, the energy eigenstates of $H^\prime$ can be written in terms of the phase $\theta_k$ of $H$ as:
\begin{eqnarray}\psi_{k\pm}^\prime (r)
&=& {1\over \sqrt{2 N}}\sum_{n=1}^N e^{ikn}\left[  \phi\left( r- n+{1\over 4}\right) \right. \nonumber\\ & & \qquad\qquad \left.
\pm e^{+i\theta_k-ik/2}\phi \left(r- n-{1\over 4}\right) \right]  \\
&=&  {1\over \sqrt{2 N}} \sum_{n=1}^N    e^{i(kn +\theta_k+k/2   )}\left[  e^{-i\theta_k-ik/2} \phi\left( r- n+{1\over 4}\right) \right. \nonumber\\ & & \qquad\qquad \left.
\pm\phi \left(r- n-{1\over 4}\right) \right]  .
\end{eqnarray}
Now shift the origin by $1\over 2$ unit, $r\to r-{1\over 2}$, and shift the summation index $n\to n-1$ in the second sum. After a few steps of algebra, one finds
that the new wavefunctions are related to the old ones by:
\begin{equation} \psi^\prime_{k\pm} (r) = \pm e^{i(\theta_k -{{3k}\over 2})} \psi_{k\pm }(r). \end{equation}  The phases linear in $k$ come from the shifts in origin for $r$ and $n$. The $A$ and $B$ wavefunctions are now:
\begin{eqnarray}
\psi_A^\prime (r)&=&{1\over \sqrt{2N}}\sum_ke^{-ikn_0}e^{i(\theta_k -{{3k}\over {2}}) )} \left( \psi_{k+} -\psi_{k-}\right) \\
\psi_B^\prime (r)&=&{1\over \sqrt{2N}}\sum_ke^{-ik(n_0+2)} \left( \psi_{k+} +\psi_{k-}\right) .
\end{eqnarray}
Aside from the extra phase factors, it can be seen that the change of winding number as the Hamiltonian changes from $H$ to $H^\prime$ has effectively converted the $A$-type wavefunctions into $B$-type
wavefunctions, and vice-versa.

\begin{figure}
\centering
\includegraphics[totalheight=1.8in]{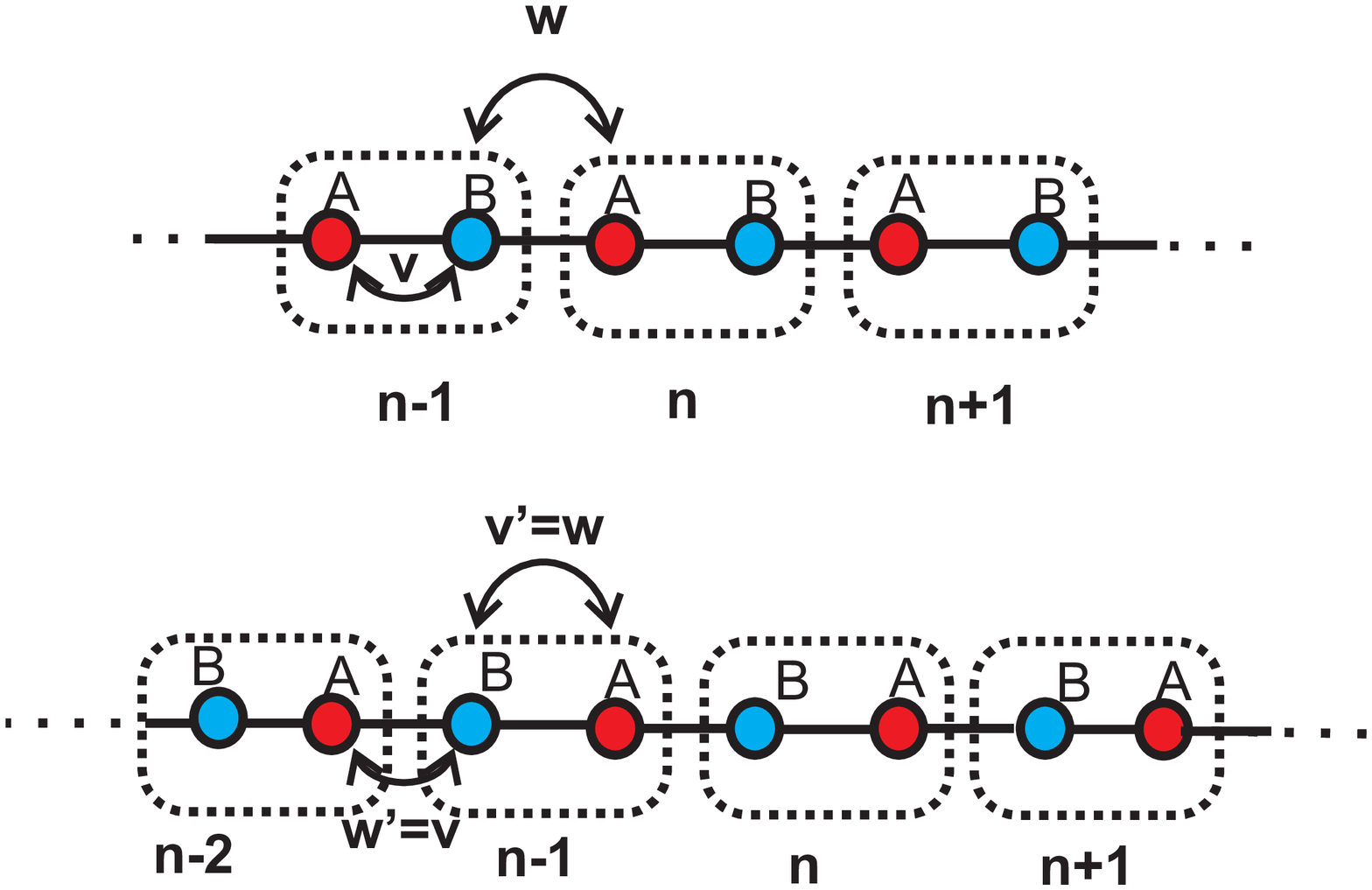}
\caption{Interchanging $v$ and $w$ amounts to shifting the units cells by a distance of $1\over  2$ units, which effectively reverses the roles of $A$ and $B$ subsites. The top and bottom figures represent the two cases. }\label{shiftfig}
\end{figure}

The interchange of $v$ and $w$ essentially redefines the unit cells, shifting each cell by one-half unit. This interchanges the roles of the $A$ and $B$
subsites, as shown in Fig. \ref{shiftfig}. Moreover, as would be expected from a topological transition, the change from $A$ being to the left of $B$ within the
unit cell to having $A$ on the right is a \emph{discrete} change,  and this change must be carried out globally on the entire system.

In the case we consider, $v^\prime =w$ and $w^\prime =v$, it should be noted that the energy eigenvalues are the same on both sides of the boundary: $E_k=E_k^\prime$, so the reflection is not due to any mismatch of energy levels. In fact, the total energy of the $A$ and $B$ modes vanishes on both sides. For example, \begin{eqnarray}E_A&=& \langle \psi_A|\hat H|\psi_A\rangle \\
&=& {1\over {2N}}\sum_{k,k^\prime}e^{-i(k-k^\prime )n_0} \\
& & \quad \times \left[ +E_k\left( \langle \psi_{k^\prime+}|\psi_{k+}\rangle +\langle \psi_{k^\prime -}|\psi_{k+}\rangle \right)\right. \\
& & \quad \quad \left.  -E_k\left( \langle \psi_{k^\prime+}|\psi_{k-}\rangle +\langle \psi_{k^\prime-}|\psi_{k-}\rangle \right)   \right] \\
&=&0 . \end{eqnarray} The vanishing of the energies makes intuitive sense: each $A$ or $B$ state is an equal superposition of eigenstates from the upper and lower bands. Since the energies of the two bands are negatives of each other, the total energy must be zero.

\begin{figure}
\centering
\includegraphics[totalheight=1.6in]{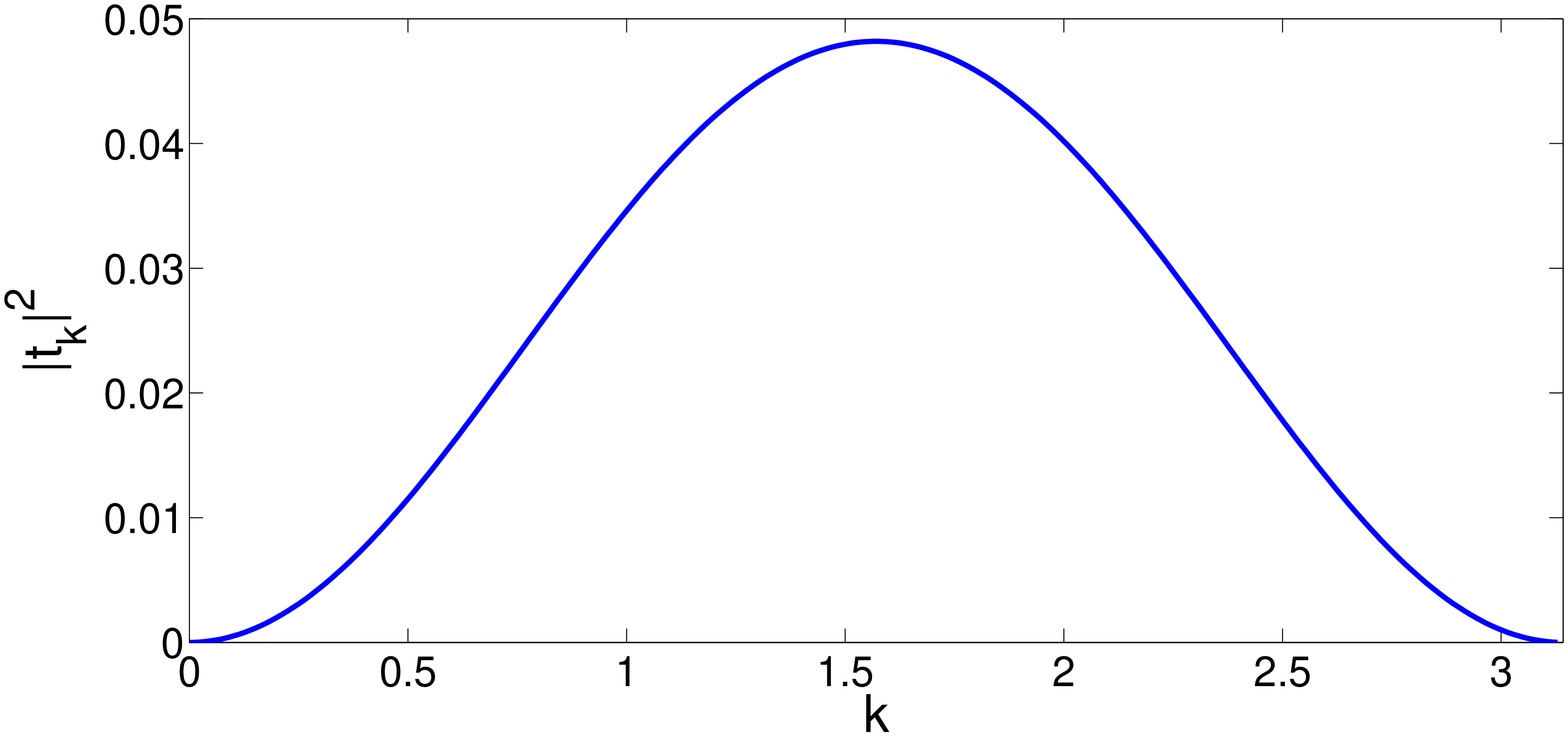}
\caption{The transmission probability $|t_k|^2$ between the regions of different winding number always peaks at $k={\pi\over 2}$, dropping to $0$ at $k=0,\pi$.
The plot here is for $N=500$, $v=0.1$, and $w=0.9$. }\label{transfig}
\end{figure}

\begin{figure}
\centering
\includegraphics[totalheight=1.6in]{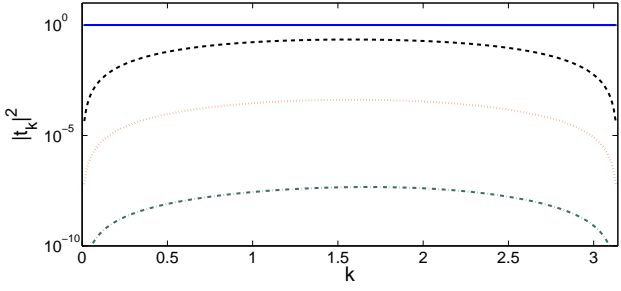}
\caption{Logarithmic plot of transmission probability $|t_k|^2$ for several values of $v$ and $w$.  The transmission probability is constant at
$|t_k|^2=1.0$ for $v=w$, but drops as $|v-w|$ increases. Increasing $|v-w|$ increases the value of $|\theta_k|$ at each nonzero $k$, leading to a larger phase
shift at the boundary. The values plotted here are $(v,w)= (.5.5)$ (solid blue), $(.2,.8)$ (dashed black), $(.01,.99)$ (dotted red), and $(.001,.999)$
(dash-dot green). As $|v-w|\to 1$, the peak transmission $|t_k|^2\to 0$.}\label{logtransfig}
\end{figure}

A more quantitative examination can be made of the transition across the boundary. We take the boundary to occur at the $n=0$ cell, with the roles of $v$ and $w$
reversing once the $B$ subsite of that cell is crossed. Only right-moving modes can cross from the left side to the right, so consider a positive-energy
right-moving mode ($k>0$) encountering the boundary; some of the amplitude can reflect to the left, some may be transmitted to the right. In addition, a
localized edge state can be built up around the boundary. So one may consider a state of the form
\begin{equation}|\Psi\rangle =|\psi_{k+}\rangle +r_k|\psi_{-k,+}\rangle +t_k|\psi^\prime_{k+}\rangle +|\psi_{e,k} \rangle  ,\end{equation}
where the terms on the right represent the incident, reflected, transmitted, and edge states. (A similar state can be constructed using a right-moving negative
energy state with $k<0$; the results will be similar.) This state must satisfy the eigenvalue equation \begin{equation}\hat H|\Psi\rangle
=E_k|\Psi\rangle.\label{eigeq}
\end{equation} Eq. \ref{eigeq} can be solved exactly: the solutions for $a_0$, $b_0$, $a_1$, $b_1$, as well as for the reflection and transmission amplitudes
$r_k$ and $t_k$, are given in the appendix. For given values of $v$ and $w$, the transmission probability at a fixed $k$ value, $|t_k|$, peaks at $k={\pi\over 2}$,
dropping to zero at $k=0,\pi$ (Fig. \ref{transfig}). When the two hopping amplitudes are equal, $v=w$, the transmission is $100\%$, as would be expected, since
the two sides of the boundary are identical at this value. However, as the difference $|v-w|$ increases, the peak transmission drops (Fig. \ref{logtransfig}).

We see then that by choosing $w$ close to $0$ and $v$ close to $1$, or vice versa, we can make the transmission of the wavefunction into the second region
arbitrarily small, even though the energy levels are identical on both sides. A figure of merit might be taken to be \begin{equation}T_{max} \equiv max_k\left( |t_k|^2\right) ,\end{equation} the fixed-$k$ transition probability, maximized over all $k$ values. Then, for example, if it is desired to keep $T_{max}\leq 10^{-3}$, this can be accomplished by arranging to have $\left| {{v-w}\over {v+w}}\right|>.96$

\section{Conclusion}\label{concludesection}

In this paper, we have constructed explicit expressions for the position-space wavefunctions of the SSH model in the case of an initially localized particle, and
examined what happens to them when the propagating wavefunction encounters a change in system parameters that discontinuously alters the system's winding number.
Any modes hitting the interface between regions exhibits some reflection backward. Therefore transmissions across the boundary are suppressed, and if $v$ and $w$ are well-chosen
they can be made arbitrarily small. It is well-known that localized edge states existing at boundaries between regions with different values of a topological
invariant enjoy a form of protection against perturbations. The results here imply that under some conditions a weaker form of protection can be made to extend
to states in the bulk regions: if a severe external perturbation to the Hamiltonian causes a change in winding number in some region, the wavefunction resists
entering that region, and tends to stay in the unperturbed region of original winding number. Since this happens due to the change in a topological quantum
number, it could be referred to as ``topologically-assisted suppression of transitions'' of the bulk wavefunction. This effect has obvious applications, since it can be used to
protect quantum information encoded in bulk wavefunctions against environment-induced errors. Such applications will be examined in detail elsewhere.

\section*{Acknowledgements} This research was supported by the National Science Foundation EFRI-ACQUIRE Grant No. ECCS-1640968, AFOSR Grant No. FA9550-18-1-0056, and by the
Northrop Grumman NG Next.

\appendix*
\section{Exact Solution of Wavefunction Across Topological Boundary}

\begin{figure}
\centering
\includegraphics[totalheight=2.0in]{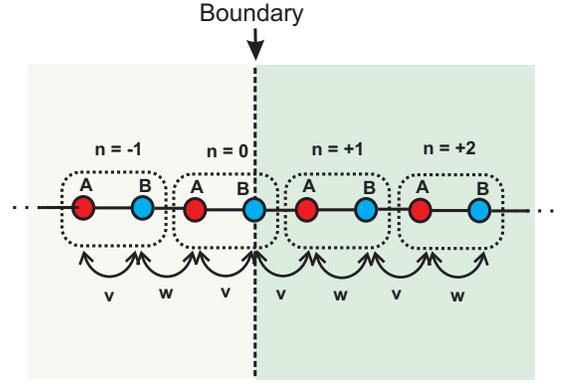}
\caption{The boundary between distinct topological regions is taken to pass through the $B$ subcell of site $n=0$. When the boundary is crossed, the roles of $v$ and $w$ are interchanged.}
\label{boundaryfig}
\end{figure}

Consider a right-moving state coming from the left and hitting the boundary between the two regions of different winding number. We will consider only
states on the upper energy band; the negative energy band is similar. The full state of the system can be written in the form
\begin{equation}|\Psi\rangle =|\psi_{k+}\rangle +r_k|\psi_{-k,+}\rangle +t_k|\psi^\prime_{k+}\rangle +|\psi_{e,k} \rangle  \label{app1},\end{equation}
for $k>0$ (right-moving incident wave), where the terms on the left represent, respectively, the incident, reflected, and transmitted state, as well as a localized edge state
$|\psi_{e,k}\rangle  $. We take the boundary to pass through the $B$ subcell of the $n=0$ site. Referring to Fig. \ref{boundaryfig}, the transition between the
two asymptotic solutions takes place over the span of the $n=0$ and $n=1$ sites, so the edge state may be expressed as
\begin{equation}|\psi_{e,k}\rangle ={1\over \sqrt{2N}}\left( a_0 |A,0\rangle +b_0|B,0\rangle + a_1 |A,1\rangle +b_1|B,1\rangle\right) . \label{app2}\end{equation}

Using Eqs. \ref{psiAA} and \ref{psiBB}, the state may be written as  \begin{eqnarray}|\Psi\rangle &=& {1\over \sqrt{2N}}\left\{\sum_{-N+1}^{-1}\left[ e^{ikn}\left( |A,n\rangle +e^{-i(\theta_k-{k/2})}|B,n\rangle\right)\right.\right.\nonumber \\ & & \quad+\left. r_ke^{-ikn}\left( |A,n\rangle +e^{i(\theta_k-k/2)}|B,n\rangle\right)\right] \\
& & \quad  +t_k\sum_{n=2}^N e^{ikn}\left( |A,n\rangle +e^{i(\theta_k +k/2)}|B,n\rangle \right)\nonumber \\
& & + \left. a_0 |A,0\rangle +b_0|B,0\rangle + a_1 |A,1\rangle +b_1|B,1\rangle  \right\} .  \nonumber \end{eqnarray}
Here, use has been made of the fact that the sign of $\theta_k$ changes as the boundary is crossed.

The Hamiltonian is of the form of Eq. \ref{coordfig} to the left of the boundary, while on the right it is of the same form with $v$ and $w$ interchanged.
Keeping mind that the interchange of $v$ and $w$ also flips the sign of $\theta_k$, this gives
\begin{eqnarray} &\hat H|\Psi\rangle& = {1\over \sqrt{2N}}\left\{\sum_{-N+1}^{-1}\left[ \left( e^{ikn}+r_ke^{-ikn}\right)\right.\right. \\
& & \qquad \qquad\qquad \times\left( v|B,n\rangle +w|B,n-1\rangle\right)\nonumber \\ &&  +\left( e^{ikn}e^{-i(\theta_k-{k/2})}+r_ke^{-ikn}e^{i(\theta_k-{k/2})}\right) \nonumber \\ && \left.
\qquad \qquad\qquad \times \left( v|A,n\rangle +w|A,n+1\rangle\right)\right] \nonumber \\
& &   +t_k\sum_{n=2}^N e^{ikn} \left[ \left( v|B,n-1\rangle +w|B,n\rangle \right) \right.  \nonumber \\
& & \left. +e^{i(\theta_k+k/2)}\left( v|A,n+1\rangle +w|A,n\rangle \right)\right] \nonumber \\
& &  +a_0\left( v|B,0\rangle +w|B,-1\rangle \right) + b_0\left( v|A,0\rangle +v|A,1\rangle \right) \nonumber \\
& & \left. +a_1\left( v|B,0\rangle +w|B,1\rangle \right) + b_1\left( w|A,1\rangle +v|A,2\rangle \right)  \right\} . \nonumber \end{eqnarray}

The state must satisfy $\hat H |\Psi\rangle =E_k|\Psi \rangle$. Equating terms of the same kind ($|A,1\rangle$, $|B,0\rangle$, etc.) on each side leads to a set
of equations that can be combined into a matrix equation of the form  \begin{equation}M\cdot V=W,  \end{equation}
where \begin{equation}M=\left(\begin{array}{cccccc} 0 & v & -E_k & w & 0& 0\\
v & -E_k & v & 0 & 0 & 0 \\
0 & 0 & w & -E_k & 0 & vy^4\\
-E_k & v & 0 & 0 & {{wy}\over x} & 0 \\
w & 0 & 0 & 0 & y(yv-{{E_k}\over x}) & 0 \\
0 & 0 & 0 & {v\over {y^{2}}} & 0 & y({{wy^2}\over x}-E_ky),
\end{array} \right)\end{equation}
and  \begin{equation}V=\left( \begin{array}{c}a_0\\ b_0 \\ a_1 \\ b_1\\ r_k \\ t_k \end{array}\right) ,\qquad
W=\left( \begin{array}{c}0\\ 0 \\ 0 \\ -wx/y\\ {{(E_k x-v/y)}\over y} \\ 0 \end{array}\right) .
\end{equation} Here, we have defined $x=e^{-i\theta_k}$ and $y=e^{ik/2}$.
Making the further definition
\begin{eqnarray}D^{-1}&=&y^2((E^6 - v^4w^2 + E^2(v^2 + w^2)(2v^2
+ w^2)\nonumber \\ & &  - E^4(3v^2 + 2w^2))x + (E^2 - v^2)vx^2)y\nonumber \\
& & - E(E^2 - 2v^2 -
w^2)((E^2 - w^2)w\\ & &   +
vw(E^4 + v^2w^2 - E^2(2v^2 + w^2))xy^2)\nonumber ,\end{eqnarray}
then the solutions of these equations are given by
\begin{eqnarray}
a_0&=& -D\left( vw\left( (E_k^4 + v^4 - E_k^2(2v^2 + w^2))x\right. \right. \\
    & & +\left. \left. E_kw(-E_k^2 + v^2 + w^2)y\right)( x^2y^2-1)\right)\nonumber \\
b_0&=& D\left(v^2w\left(-E_k^3x + E_k(v^2 + w^2)x \right.\right.\\ & & \left.\left. + E_k^2wy
    - w^3y\right)( x^2y^2-1)\right)\nonumber  \\
a_1 &=& D(v^3w(-E_k^2x + v^2x + E_kwy)( x^2y^2-1) )\\
b_1 &=& D( v^3w^2(-E_kx + wy)( x^2y^2-1) ) \\
r_k &=& Dy^{-1}\left( x\left(E_kv(v^2-E_k^2)(-E_k^2 + 2v^2 + w^2)x\right.\right. \nonumber \\ & &  -
   \left( E_k^6x^2 + v^3w^2(w - vx^2)\right. \label{appr}\\ & & +
    E_k^2(2v^2 + w^2)(-vw + (v^2 + w^2)x^2) \nonumber \\ & & +
     \left. E_k^4(vw - (3v^2 + 2w^2)x^2)\right) y \nonumber \\ & & +\left.\left.
    E_k w(E_k^2 - w^2)(E_k^2 - 2v^2 - w^2)xy^2\right) \right) \nonumber \\
t_k &=& Dy^{-4}\left( v^4w^2x\left( 1 - x^2y^2\right)\right)   .\label{appt}
\end{eqnarray}
Substituting these formulas into Eqs. \ref{app1}-\ref{app2} gives an exact solution to the eigenvalue problem. The reflection and transmission coefficients of
Eqs. \ref{appr} and \ref{appt} were used to construct the plots of Figs. \ref{transfig} and \ref{logtransfig}.

%


\begin{thebibliography}{99}


\bibitem{su} W. P. Su, J. R.  Schrieffer, and A. J. Heeger, \emph{Phys. Rev.
    B} \textbf{22}, 2099 (1980).

\bibitem{hasan} M. Z. Hasan, C. L. Kane, \emph{Rev. Mod. Phys.} \textbf{82}, 3045 (2010)

\bibitem{asboth} J. K. Asb\'{o}th, L. Oroszl\'{a}ny, A. P. P\'{a}lyi, \emph{A Short Course on Topological Insulators: Band Structure and Edge States in
    One and Two Dimensions} (Springer, Heidelberg, 2017 )


\bibitem{kitagawatop} T. Kitagawa, M. S. Rudner, E. Berg, and E. Demler, \emph{Phys. Rev. A} \textbf{82}, 033429 (2010)

\bibitem{broome} M. A. Broome, A. Fedrizzi, B. P. Lanyon, I. Kassal,
    A. Aspuru-Guzik, A. G. White, \emph{Phys. Rev. Lett.} \textbf{104},
    153602 (2010)

\bibitem{kit1} T. Kitagawa, M. S. Rudner, E. Berg, E. Demler, \emph{Phys.
    Rev. A} \textbf{82}, 033429 (2010)

\bibitem{kit2} T. Kitagawa, E. Berg, M. Rudner, and E. Demler,   
\emph{Phys. Rev. B.} \textbf{82}, 235114 (2010)

\bibitem{kit3}  T. Kitagawa, M. A. Broome, A. Fedrizzi, M. S. Rudner, E. Berg,  I. Kassal,  A. Aspuru-Guzik,  E. Demler, and A. G. White, \emph{Nature
    Comm}. {\bf 3}, 882 (2012)

\bibitem{taras} B. Tarasinski, J. K. Asb\'oth, and J. P. Dahlhaus,
    \emph{Phys. Rev. A} \textbf{89}, 042327 (2014)

\bibitem{cardano} F. Cardano, A. D'Errico, A. Dauphin, M. Maffei, B. Piccirillo, C. de Lisio,  G. De Filippis, V. Cataudella,  E. Santamato, L. Marrucci,
    M. Lewenstein, P. Massignan, \emph{Nat. Comm.} \textbf{8}, 15516 (2017)

\bibitem{sim3} D. S. Simon, C. A. Fitzpatrick, S. Osawa, and A. V. Sergienko,
    \emph{Phys. Rev. A} \textbf{96}, 013858 (2017)

\bibitem{sim2} D. S. Simon, C. A. Fitzpatrick, S. Osawa, and A. V. Sergienko,
    \emph{Phys. Rev. A} \textbf{95}, 042109 (2017)

\bibitem{kempe} J. Kempe, \emph{Contemp. Phys.} \textbf{44}, 307 (2003) 

\bibitem{amba} A. Ambainis, 
\emph{Int. J. Quant. Inf.} \textbf{1}, 507 (2003)

\bibitem{portugal} R. Portugal,  \emph{Quantum Walks and Search Algorithms} (Springer, Berlin, 2013)

\bibitem{jackiw} R. Jackiw and C. Rebbi,   
\emph{Phys. Rev. D} \textbf{13}, 3398 (1976).

\bibitem{wan} G. H. Wannier, 
\emph{Phys. Rev. } \textbf{52}, 191 (1937).

\bibitem{kohn} W. Kohn, 
\emph{Phys. Rev.} \textbf{115}, 809 (1959).

\bibitem{cloiz} J. des Cloizeaux, 
\emph{Phys. Rev.} \textbf{129}, 554 (1963).




\end{thebibliography}
\end{document}